\documentclass[pdflatex,sn-mathphys-num,iicol]{sn-jnl}

\usepackage{graphicx}%
\usepackage{multirow}%
\usepackage{amsmath,amssymb,amsfonts}%
\usepackage{amsthm}%
\usepackage{mathrsfs}%
\usepackage[title]{appendix}%
\usepackage{xcolor}%
\usepackage{textcomp}%
\usepackage{manyfoot}%
\usepackage{booktabs}%
\usepackage{algorithm}%
\usepackage{algorithmicx}%
\usepackage{algpseudocode}%
\usepackage{listings}
\usepackage{subcaption}%
\usepackage[acronym]{glossaries}
\usepackage{lipsum} 
\usepackage{rotating}
\usepackage[utf8]{inputenc}
\usepackage{cuted}

\usepackage[utf8]{inputenc}  
\usepackage[T1]{fontenc}      
\usepackage{textcomp}         

\loadglsentries{gloss.tex}

\theoremstyle{thmstyleone}

\theoremstyle{thmstyletwo}%

\theoremstyle{thmstylethree}%

\begin{document}

\title[Article Title]{Atmospheric Muon Flux Suppression at Potential New Low-Radiation Underground Physics Laboratory in Israel}

\author*[1]{\fnm{Nadav} \sur{Hargittai}}\email{nadav.hargittai@weizmann.ac.il}

\author[2]{\fnm{Igor} \sur{Zolkin}}

\author[2,3]{\fnm{Yiftah} \sur{Silver}}

\author[2]{\fnm{Yan} \sur{Benhammou}}

\author[2]{\fnm{Gilad} \sur{Mizrachi}}

\author[1]{\fnm{Hagar} \sur{Landsman}}

\author[2]{\fnm{Erez} \sur{Etzion}}

\author[1]{\fnm{Ranny} \sur{Budnik}}

\affil[1]{\orgdiv{Department of Particle Physics and Astrophysics}, \orgname{Weizmann Institute of Science}, \city{Rehovot}, \country{Israel}}

\affil[2]{\orgdiv{Raymond and Beverly Sackler School of Physics and Astronomy}, \orgname{Tel Aviv University}, \city{Tel Aviv}, \country{Israel}}

\affil[3]{\orgname{Rafael Advanced Defense Systems LTD}, \country{Israel}}

\abstract{The residual atmospheric muon flux was measured at a candidate site for a new underground, low-radiation physics laboratory beneath the \textit{Kokhav HaYarden} national park in Israel. Located inside the tunnels of a hydroelectric pumped-storage facility operating since 2024, the proposed site benefits from a vertical rock overburden of 361 meters, large potential floorspace, and easy access by road. A muon hodoscope of vertically stacked wide-area $144 \times 60 \times 1.2 $ cm$^3$ plastic scintillator plates was employed to measure the suppression in the integrated muon flux at the site as compared with above ground at sea level. Data-taking took place in mid-August of 2024 for several days and was split into South-North and West-East orientations to account for the directional acceptances due to the geometry of the detector. The suppression factor is reported at $4456 \pm 77$, expressed as $3.75 \pm 0.06 \times 10^{-6}$ cm$^{-2}$ s$^{-1}$ in absolute terms, corresponding to an effective overburden of roughly $873$ m.w.e.. A deeper location at the site may also be available, but it could not be reached at this time. Furthermore, the asymmetric topography of the mountain above and its muon shadow are clearly visible in the angular data. Finally, auxiliary environmental measurements recorded low background radon activity at $28.3 \pm 14.0$ Bq m$^{-3}$. The experimental campaign thus succeeded in demonstrating the viability of the site's working conditions for future scientific research.}


\maketitle

\section{Introduction}

\begin{figure*}[t] 
    \centering
    \includegraphics[width=0.95\linewidth]{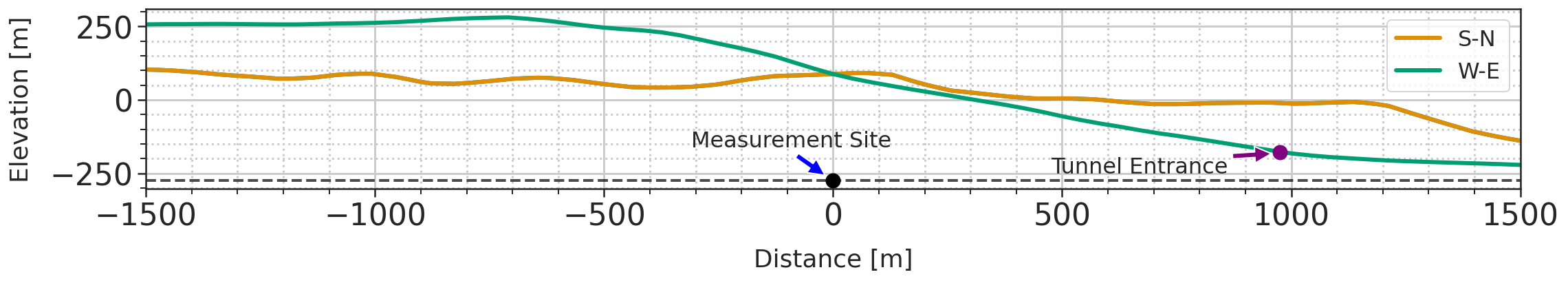}
    \caption{S-N (South-North) and West-East (W-E) profile cuts of the topography above the measurement site at KY. Elevation is above and below sea level.}
    \label{fig:geo_profiles}
\end{figure*}

Operating low-background physics experiments often requires going underground to find shielding from high levels of atmospheric muons, the most abundant high-energy component of the cosmic ray flux on the surface. Several such underground laboratories exist in the world, hosting a range of experiments in areas such as rare-event searches in dark matter direct detection \cite{DM_DD_REVIEW, DM_DD_REVIEW_2}, neutrinoless double-beta decay \cite{DOUBLE_BETA_REVIEW}, and neutrino physics \cite{BOREXINO, RENO_NEUTRINOS}. In the field of quantum information science, the need to shield superconducting qubits from ionising radiation to achieve longer coherence times and higher fidelity has also sparked interest in underground environments \cite{QM1, QM2, QM3_muonattenuation}.

Therefore, the suppressed atmospheric muon flux is a crucial parameter in evaluating potential sites. This quantity must be measured directly since accurate geological profiles for use in simulations are hard to obtain.

Preliminary work in identifying sites for a new low-radiation underground laboratory in Israel has begun by exploring one of two locations, both of which are part of pumped-storage hydroelectric projects that require boring service tunnels deep into the sides of mountains. Smaller tunnel branches, used during construction but no longer necessary thereafter, provide potentially prime real-estate. There is a precedent for taking advantage of these facilities at the YangYang underground laboratory in the Republic of Korea \cite{YANGYANG_LAB, YANGYANG_LAB2}. This work presents the first exploratory campaign performed at Kokhav HaYarden (KY) pumped storage facility \cite{KY_pumpedstorage} in the Valley of Springs regional council in Israel, at a vertical depth of $361$~m. The second pumped-storage facility, located close to Kibbutz Manara in the Upper Galille Regional Council \cite{manara}, is still under construction but promises over double the depth at approximately $750$~m.

\section{Kokhav HaYarden Site}

The \textit{Kokhav HaYarden} (KY) pumped storage facility sits squarely in between the town of Beit-She'an and the southern tip of the Sea of Galilee. It takes advantage of the height difference between the Jordan valley and the KY national park above to its west for the operations of its hydroelectric power plant. A roughly $1$~km long service tunnel, easily accessible by even large vans, extends horizontally west into the mountain from Route 90 in the valley, leading to the large engine and turbine halls, from which lesser tunnels branch out. It was at the entrance to one of these unused branches that we could perform measurements. Most of the unused branches are blocked off by a concrete wall, and access is currently limited due to increased safety requirements. However, the potential available floor space where the rest of the branch is to be commissioned is estimated to be several hundred square metres with a ceiling height of $6$~m. Furthermore, a $25\%$ increase in vertical overburden is possible deeper into the tunnel as compared with $361$~m at the measurement location. For safety reasons, we were prevented from using those deeper tunnels at this time.

\begin{figure}[h!]
    \centering
    \includegraphics[width=.95\linewidth]{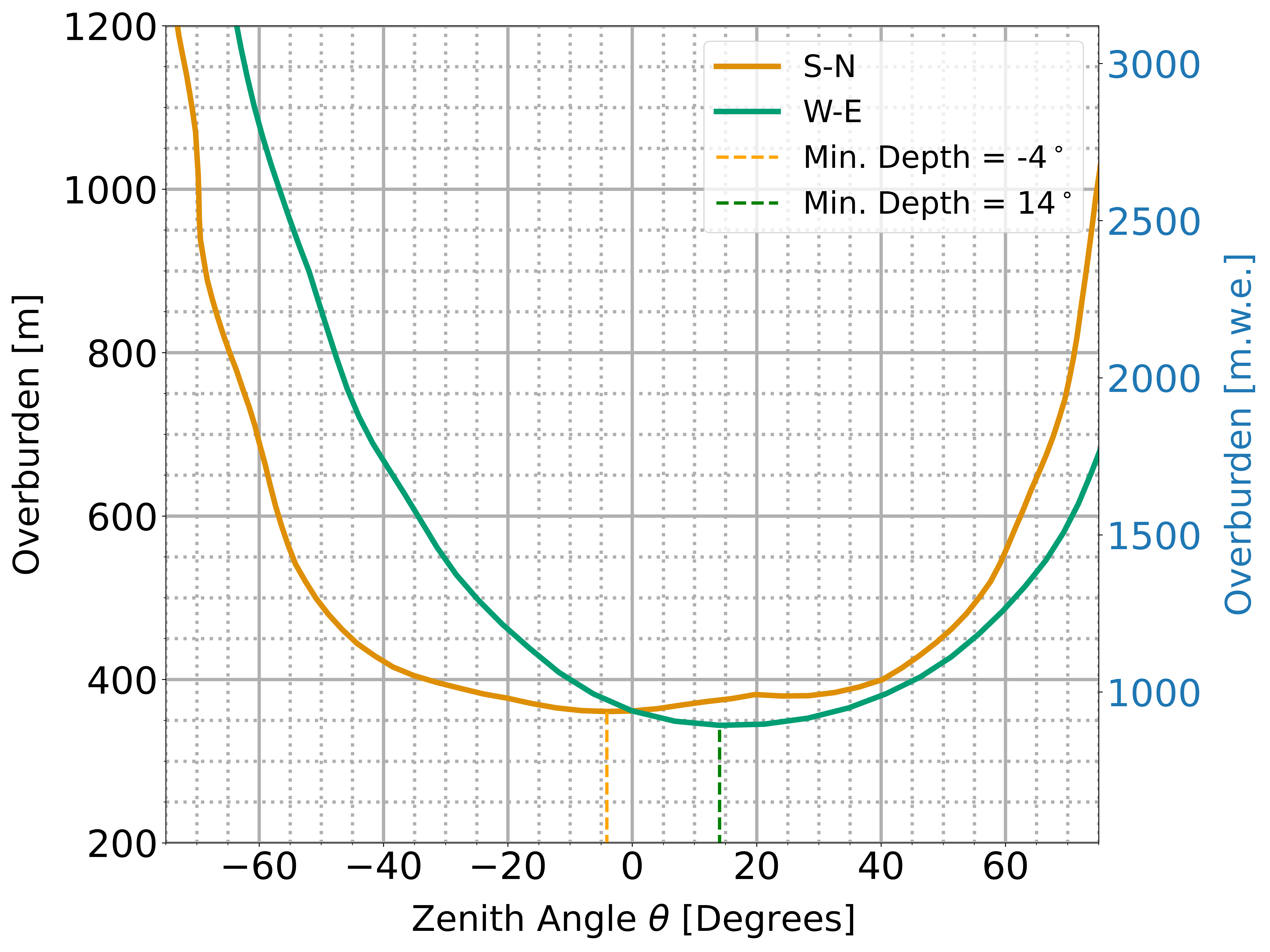}
    \caption{Slant depths, defined as the length of earth with respect to the measurement site, shown both for S-N and W-E orientations. The direction of the minimum overburden for each orientation is indicated by the dashed lines. The right axis displays the overburden in units of m.w.e., using a representative rock density value of $2.6$ g cm$^{-3}$.}
    \label{fig:geo_slants}
\end{figure}

\begin{figure}
    \centering
    \includegraphics[width=0.95\linewidth]{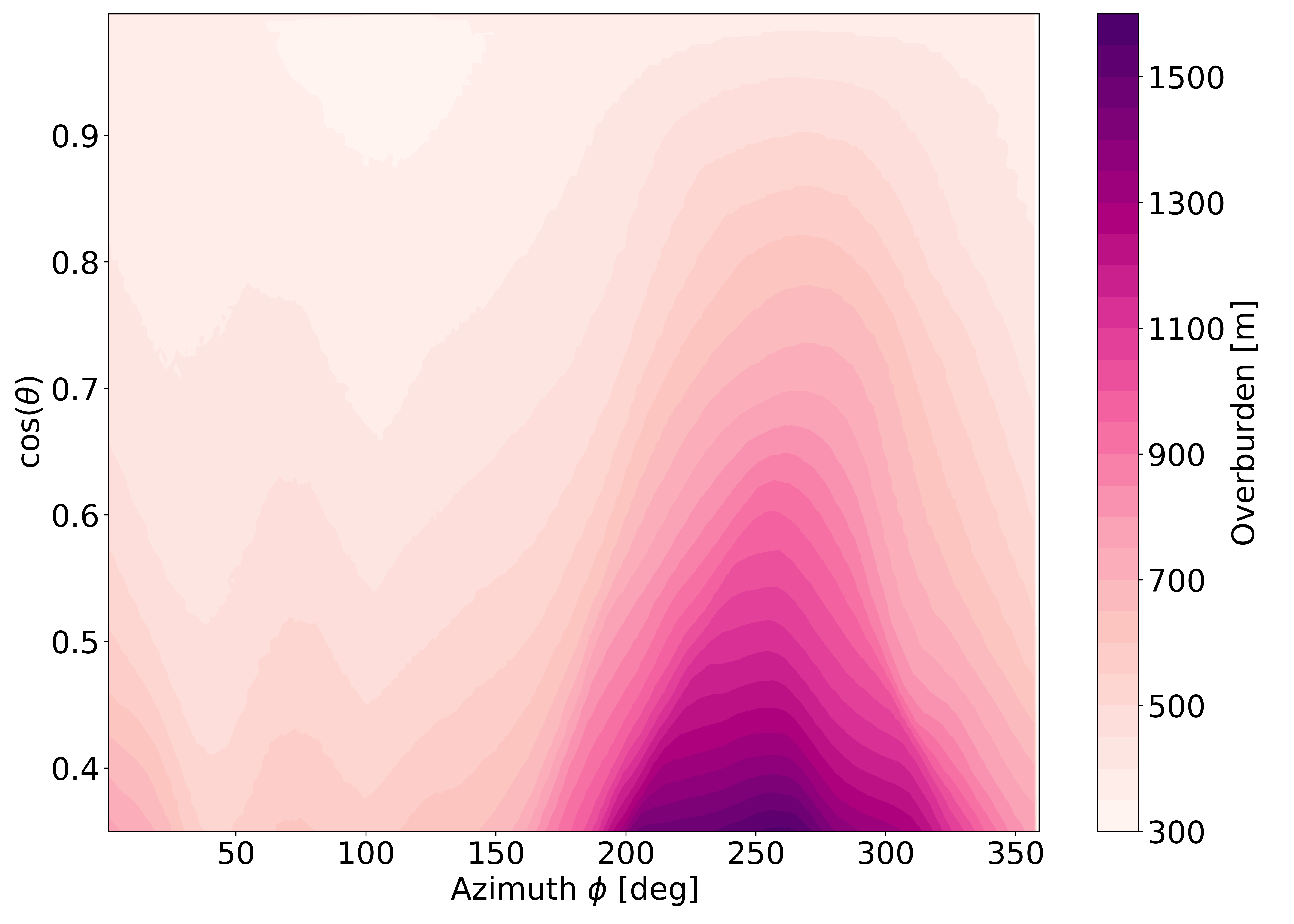}
    \caption{Full sky slant depth. The $\phi$ angle is measured from the geographical north.}
     \label{fig:slantheatmap2d}
\end{figure}

\begin{figure}
    \centering
    \includegraphics[width=0.95\linewidth]{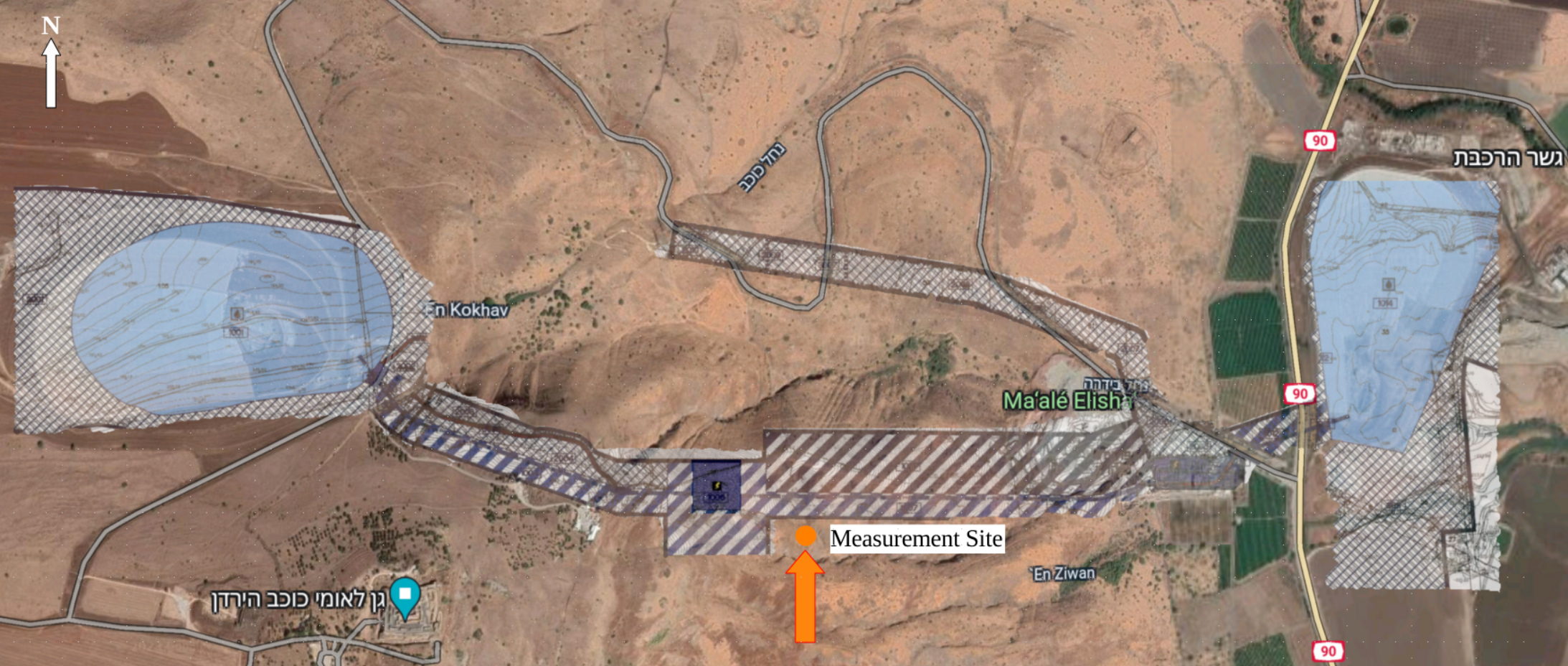}
    \caption{Satellite image with an overlay of the service tunnel map, showing the location of the measurements site \cite{googleearth_gesher}.}
    \label{fig:overlay}
\end{figure}

The South-North (S-N) and West-East (W-E) overburden profile cuts, shown in figure \ref{fig:geo_profiles}, were extracted from a surface elevation map using Google's elevation API \cite{googleelevation2025} in accordance with the separate S-N and W-E oriented data-taking campaigns. The slant depth, defined as the length of overburden at a given angle with respect to the measurement site, is shown in figure \ref{fig:geo_slants}. 
While this figure presents the S-N and W-E slant depth profiles, a full two-dimensional slant depth distribution as a function of azimuth and zenith angle is shown in figure \ref{fig:slantheatmap2d}. A simplified simulation was performed to estimate the effective vertical overburden of the underground site. It was based on the standard sea-level cosmic-ray muon flux and the continuous energy loss model for muons in matter~\cite{pdg2022}. The muon spectrum was generated using the Gaisser parametrization, and particles were propagated in angular and energy bins through the rock overburden, using the two-dimensional slant-depth map. The resulting flux attenuation corresponds to a flat overburden of approximately 860~m.w.e. in good agreement with the measured value reported here.

\begin{figure}[h]
    \centering
    \includegraphics[width=0.95\linewidth]{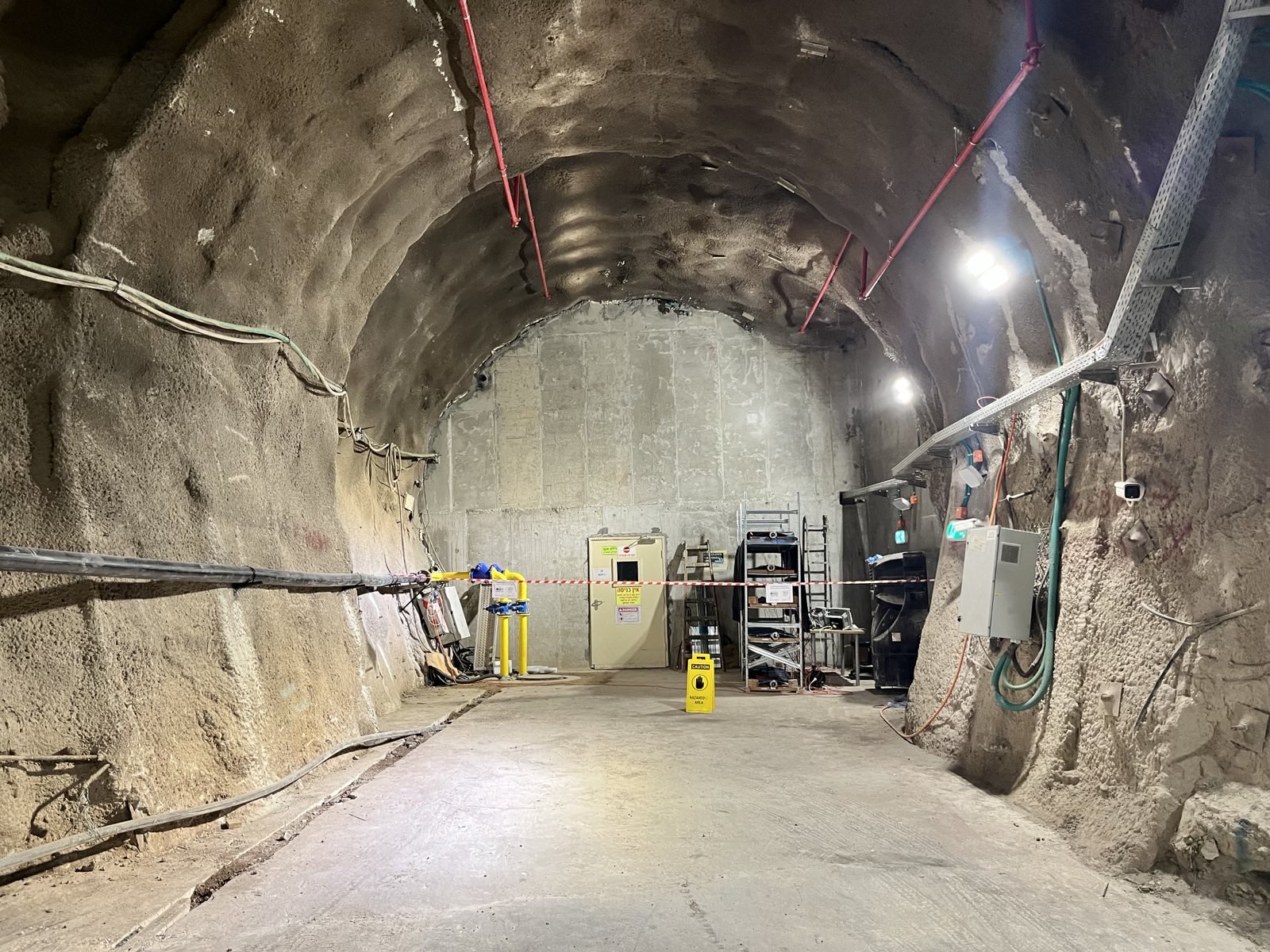}
    \caption{Photograph of the measurement site at the KY underground location, with the muon hodoscope seen on the right at a South-North orientation (the photograph is looking south). The rest of the unused tunnel and potential floorspace lie through the door behind the concrete wall. This part is currently inaccessible due to increased safety requirements in the unventilated part of the tunnel.}
    \label{fig:muon_setup_photo}
\end{figure}

The layout of the tunnel and a photograph of the measurement site are shown in figures \ref{fig:overlay} and \ref{fig:muon_setup_photo}, respectively. 

The above-ground, slightly above sea-level measurement site was located at the High Energy Laboratory at Tel-Aviv University (TAU). A summary of the two locations is shown in table \ref{tab:loc_summaries}.

\begin{table*}[t]

\centering

\caption{General properties of the measurement sites.}\label{tab:loc_summaries}%
\begin{tabular}{c|c c c c}
\toprule
Site  & Coordinates & Elevation & Vertical Overburden \\
\midrule
TAU (Tel-Aviv University)   & $32 ^ \circ 06'44''N$, $34 ^ \circ 48'23''E$  & +44 m & 0 m  \\
KY (Kokhav HaYarden) & $32 ^ \circ 35'50''N$, $35 ^ \circ 31'45''E$   & -274 m & 361 m \\
\botrule
\end{tabular}
\end{table*}

\section{Experimental Setup}

\begin{figure}[h]
    \centering
    \includegraphics[width=0.95\linewidth]{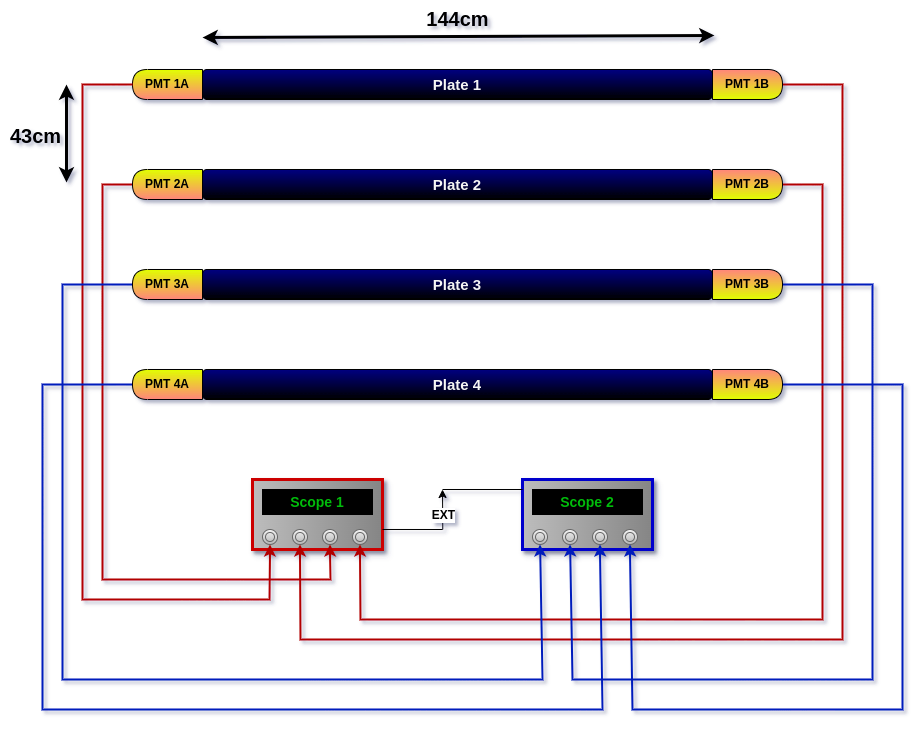}
    \caption{Schematic diagram of the muon hodoscope. The four PMTs of the top two plates are handled by one oscilloscope, which triggers the second, which handles the four PMTs on the bottom two plates.}
    \label{fig:hodoscope-Diagram}
\end{figure}

\begin{figure}[h]
    \centering
    \includegraphics[width=0.95\linewidth]{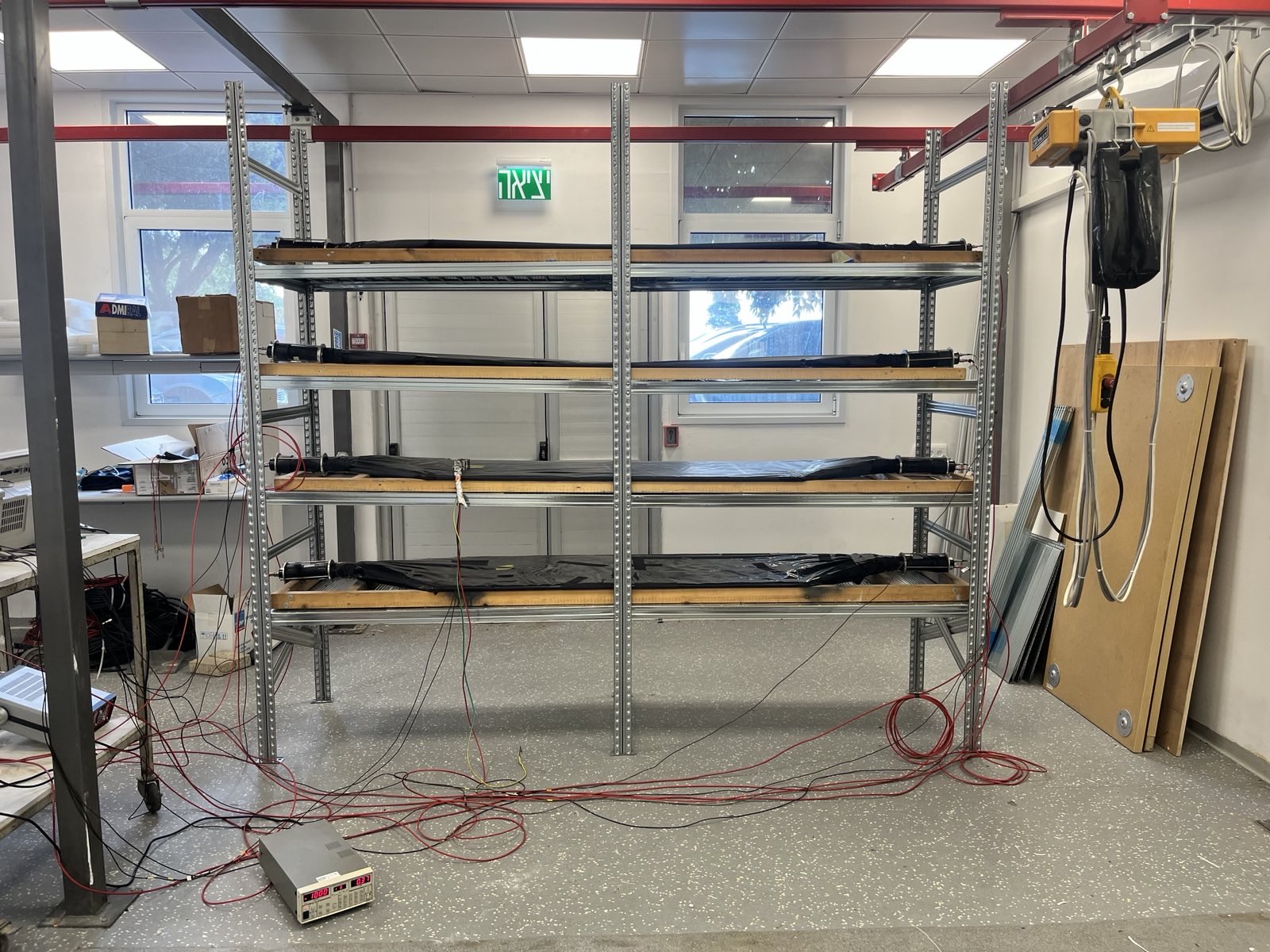}
    \caption{Photograph of the muon hodoscope above ground at Tel Aviv University, inside the High Energy Laboratory. The building has a single ground floor and is constructed with plasterboard, therefore the overburden at this site was considered to be zero.}
    \label{fig:hodoscope-photo}
\end{figure}

Measurements were performed using a muon hodoscope made up of four rectangular plastic scintillator plates \cite{HodoscopeTLV, HodoscopeTLV2} stacked vertically one above the other using a modular aluminium frame. The scintillators had an active volume of $144 \times 60 \times 1.2$ cm$^{3}$, and the stacking spacing was $43$ cm. Each one possessed two \textit{Hamamatsu R329-02} Photomultiplier Tubes (PMTs) attached to both its long ends via triangular light guides, giving a total of eight channels handled by two separate 4-channel \textit{Keysight MSO-X 4054 A} oscilloscopes. High voltage to the PMTs was provided by the \textit{CAEN DT8033N} power supply. A schematic diagram and a photograph of the setup are shown in figures \ref{fig:hodoscope-Diagram} and \ref{fig:hodoscope-photo}, respectively.

\begin{figure}
     \centering
     \begin{subfigure}[b]{0.49\textwidth}
         \centering
         \includegraphics[width=\textwidth]{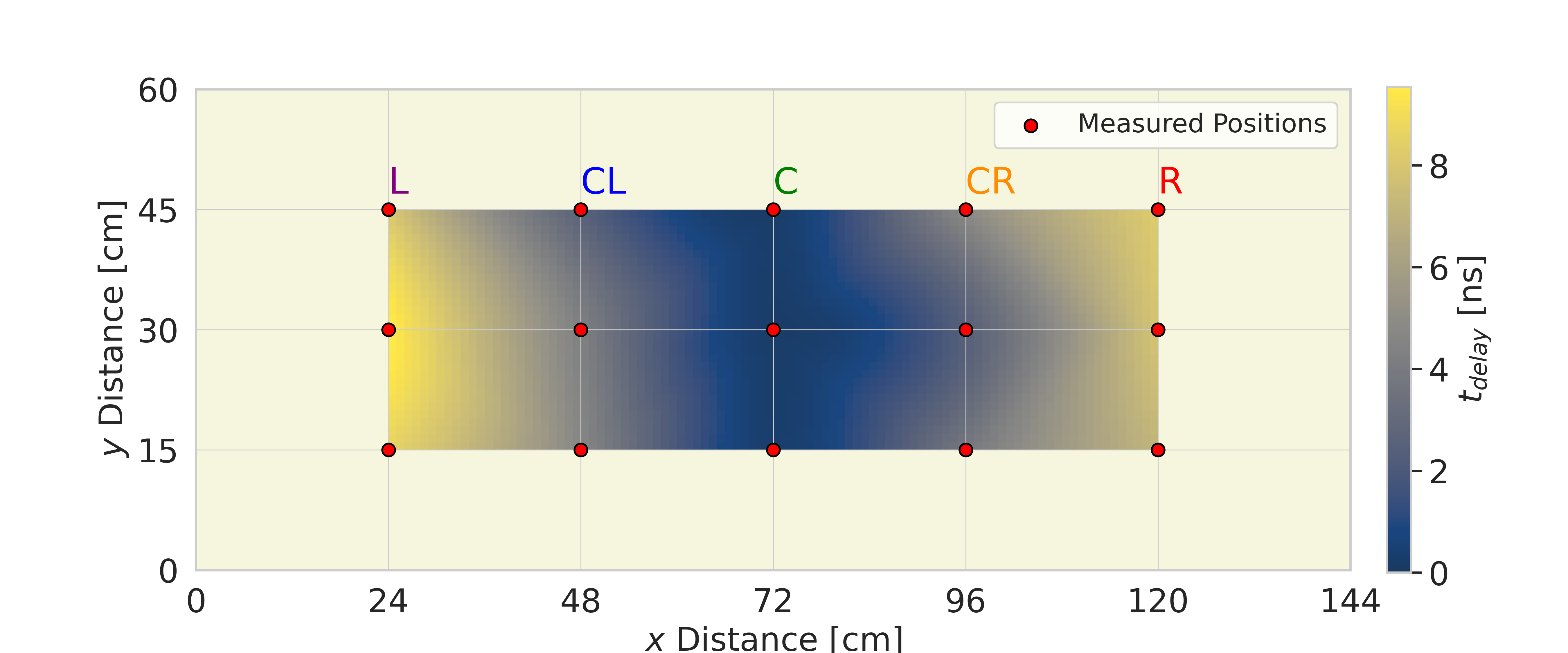}
         \caption{}
         \label{fig: calib_map}
     \end{subfigure}

     \begin{subfigure}[b]{0.49\textwidth}
         \centering
         \includegraphics[width=\textwidth]{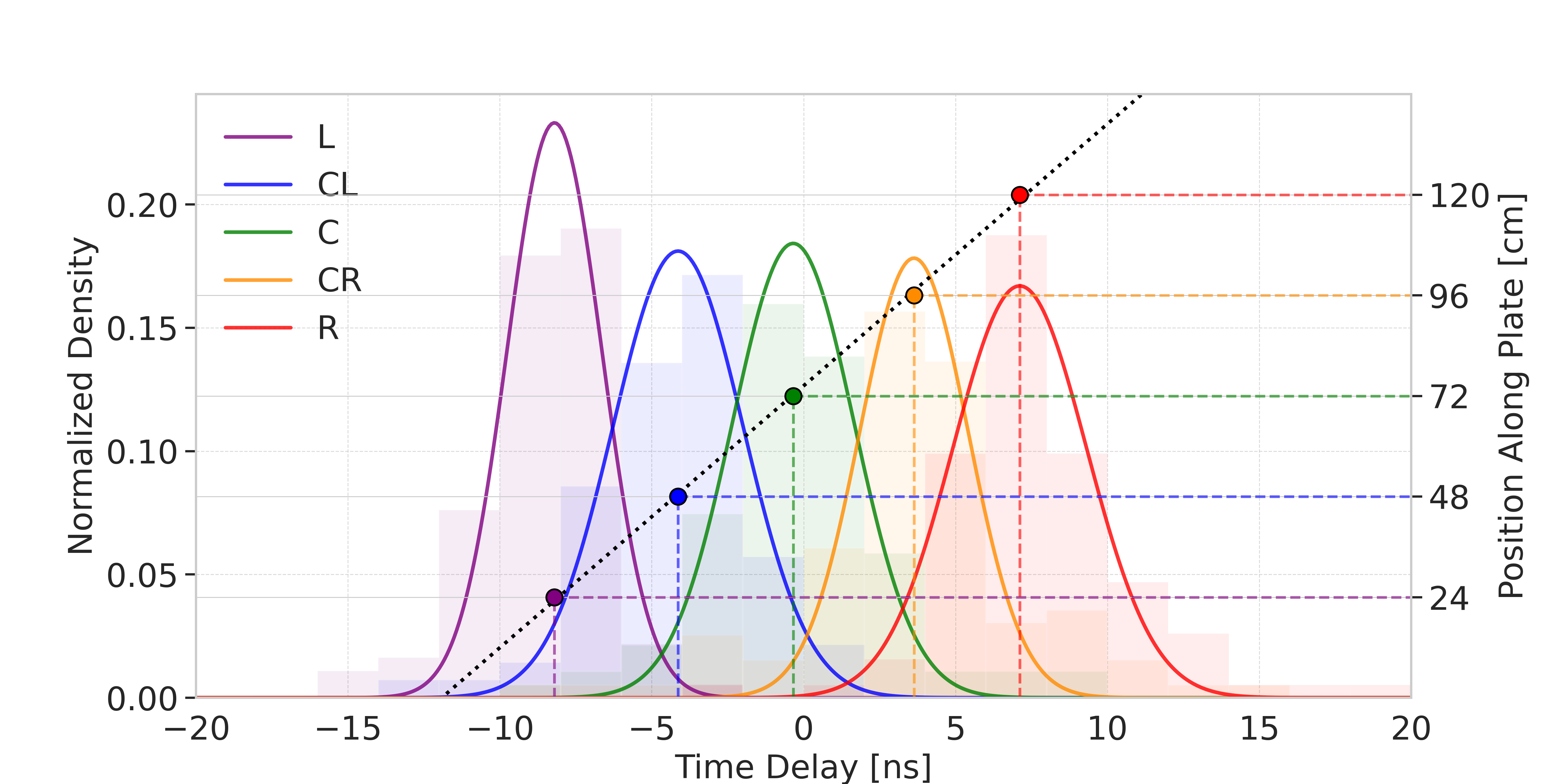}
         \caption{}
         \label{fig: calib_linear}
     \end{subfigure}
        \caption{Time-delay to position calibration. (a) Map of $t_{delay}$ shown in, the time delay in the signal arrival between two PMTs on the same plate. The red data points are the centres of the measured positions isolated with the help of a small $12 \times 25 \times 2.5$ cm$^3$ hodoscope placed on top of the plate at the specified positions. The $x$ and $y$ distances are the positions along the long and short sides of the larger scintillator plates, respectively. (b) Spread in $t_{delay}$ and linear fit to its mean at each of the measured positions of the same $x$ (the axis connecting the two PMTs): \textit{L} (Left), \textit{CL} (Centre-Left), \textit{C} (Centre), \textit{CR} (Centre-Right) and \textit{R} (Right).
        The limiting factor for assessing the errors is the size of the hodoscope used for this calibration. }
        \label{fig: calib}
\end{figure}

The trigger was set to a coincidence measurement on the four PMTs of the top two scintillators only, meaning we would obtain events containing 2-4 hits. The setup took data in both places under the same configuration to get a value for the suppression factor in the muon rate between the two locations. For angular information, track reconstruction is performed by utilising the time delay $t_{delay}$ in the arrival of a signal on each of the PMTs on the same scintillator, producing a projected one-dimensional position $x$ along the axis connecting the two PMTs. 
The timestamps are defined as the crossing times of a fixed threshold in all measurements and calibrations. 
Multiple positions of the same event are then used to recover the track and obtain the event's zenith angle of arrival $\theta$. The function $x(t_{delay})$ was calibrated by measuring the distribution of $t_{delay}$ at specific positions on the scintillator. This calibration was done by placing a second, smaller hodoscope with plate size $12 \times 25 \times 2.5$ cm$^3$ on top of our scintillators to serve as a trigger, isolate specific regions and map the $t_{delay}$ of the plates (see figure \ref{fig: calib_map}). From this map, a linear relationship is obtained, which allows us to locate events on the plates (see figure \ref{fig: calib_linear}).

The setup took data at both KY and TAU in order to compare the rates and obtain the suppression factor between the two locations. Furthermore, the setup ran at a S-N and W-E orientation at the KY site in order to account for differences in the overburden profiles. Acquisition times varied between several minutes at sea level to more than a day underground due the expected reduction in the muon flux.

\section{Results}

\begin{figure}[h]
    \centering
    \includegraphics[width=.95\linewidth]{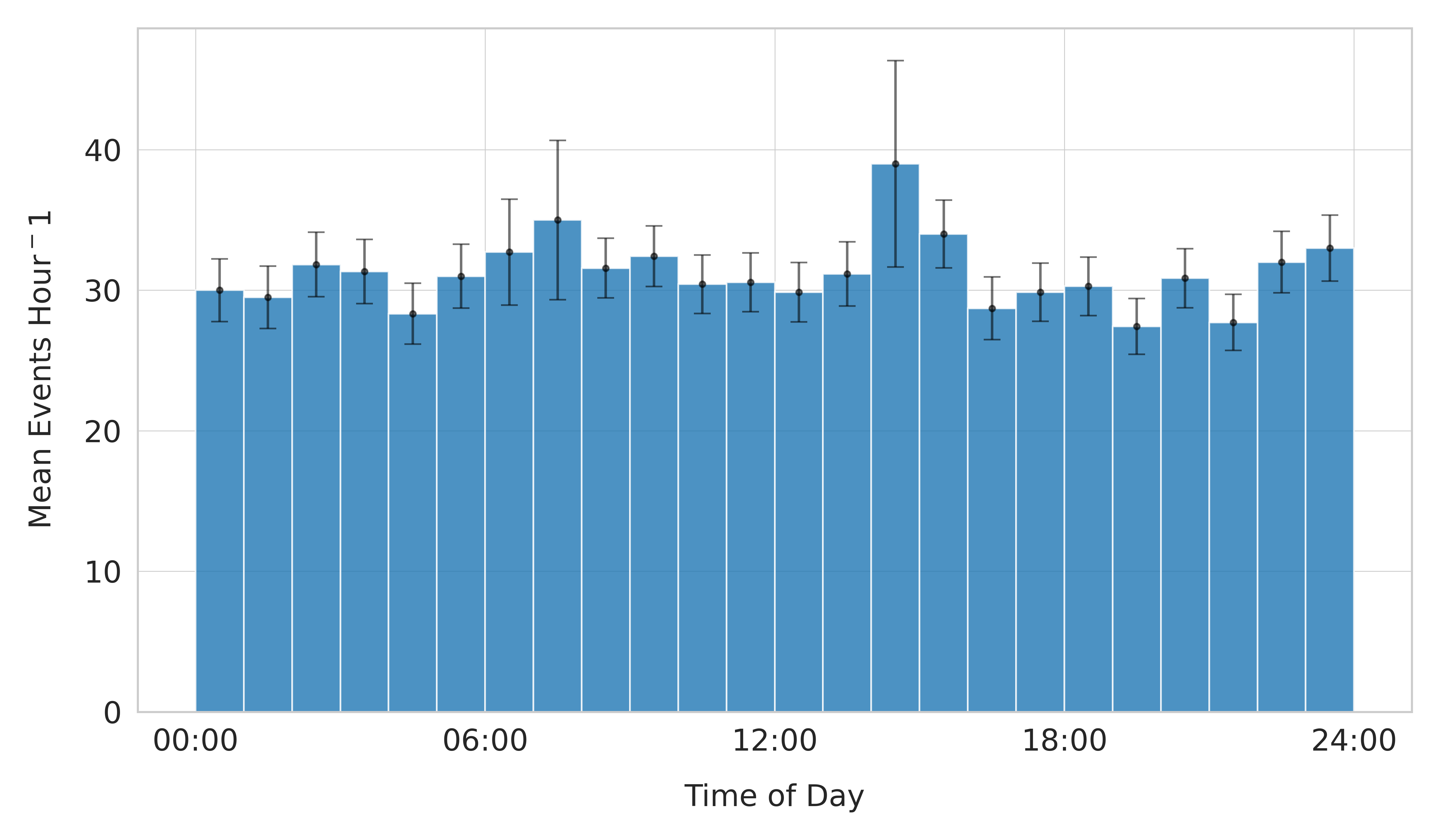}
    \caption{Muon rate as a function of time of day, represented in 1-hour-long bins, using all the measurements from both orientations at KY and statistical error bars shown.}
    \label{fig:rates}
\end{figure}

Rates were calculated by dividing the total number of triggers at each location and orientation by the total runtime and are summarised in table \ref{resultstab}. The total runtime of the KY S-N campaign is half as long as KY W-E due to an unexpected power outage during measurements, which we could not recover. The rate stood at around $0.5$ events per minute underground for the entire detector and did not change significantly throughout any single measurement. Figure \ref{fig:rates} shows the hourly rates at any given hour of the day for all the combined measurement runs at KY.

\begin{table}[h]
\centering
\caption{Summary of rates from each campaign.}\label{resultstab}%
\begin{tabular}{c|c c c}
\toprule
Site & Triggers & Time [hrs] & Rate [s$^{-1}$] \\
\midrule
TAU & 11000 & 0.08 &  $3.80 \pm 0.04 \times 10^{1}$  \\
KY S-N & 1631 & 53.1 & $8.54 \pm 0.21 \times 10^{-3}$  \\
KY W-E & 3076 & 100.1 & $8.53 \pm 0.15 \times 10^{-3}$  \\
\botrule
\end{tabular}
\end{table}

Compared with above-ground and sea-level at TAU, the suppression factor in the integrated muon flux in the KY underground site was measured at $4456 \pm 77$, utilising the ratio between the rates at TAU and the average of the two underground KY orientations. Using $1.67 \times 10^{-2}$ cm$^{-2}$ s$^{-1}$ as the known value for the muon flux at sea level \cite{review_of_particle_physics} (and neglecting the small upward-deviation at 44~m), we report an integrated muon flux of $3.75 \pm 0.06 \times 10^{-6}$ cm$^{-2}$ s$^{-1}$ at the KY underground site. We report only the statistical error here. 
Systematic uncertainties, including the detector efficiency and acceptance, are cancelled out as they are constant between every run. Multiplicity, as in two or more muons arriving at the detector but registering as a single event, is negligible for stochastic arrival of the muons. Correlated muons that trigger the detectors simultaneously are expected to be more abundant at TAU, meaning their inclusion can only increase the inferred shielding, i.e. decrease the muon flux at KY.

From our measured value of the integrated flux, we can calculate the equivalent vertical depth relative to a flat overburden using the differential muon intensity function from \cite{muon_flux_model}

\begin{equation}\label{depth_eq}
    I_\mu(h_0) = 68 \times 10^{-6} \cdot e^{-\frac{h_0}{285}} + 2 \times 10^{-6} \cdot e^{-\frac{h_0}{698}}
\end{equation}
and obtain $h_0 \sim 873$ m.w.e.. We note that equation \ref{depth_eq} is reported to be appropriate in the range 1 - 10 km.w.e..

\begin{figure}[h]
    \centering
    \includegraphics[width=.95\linewidth]{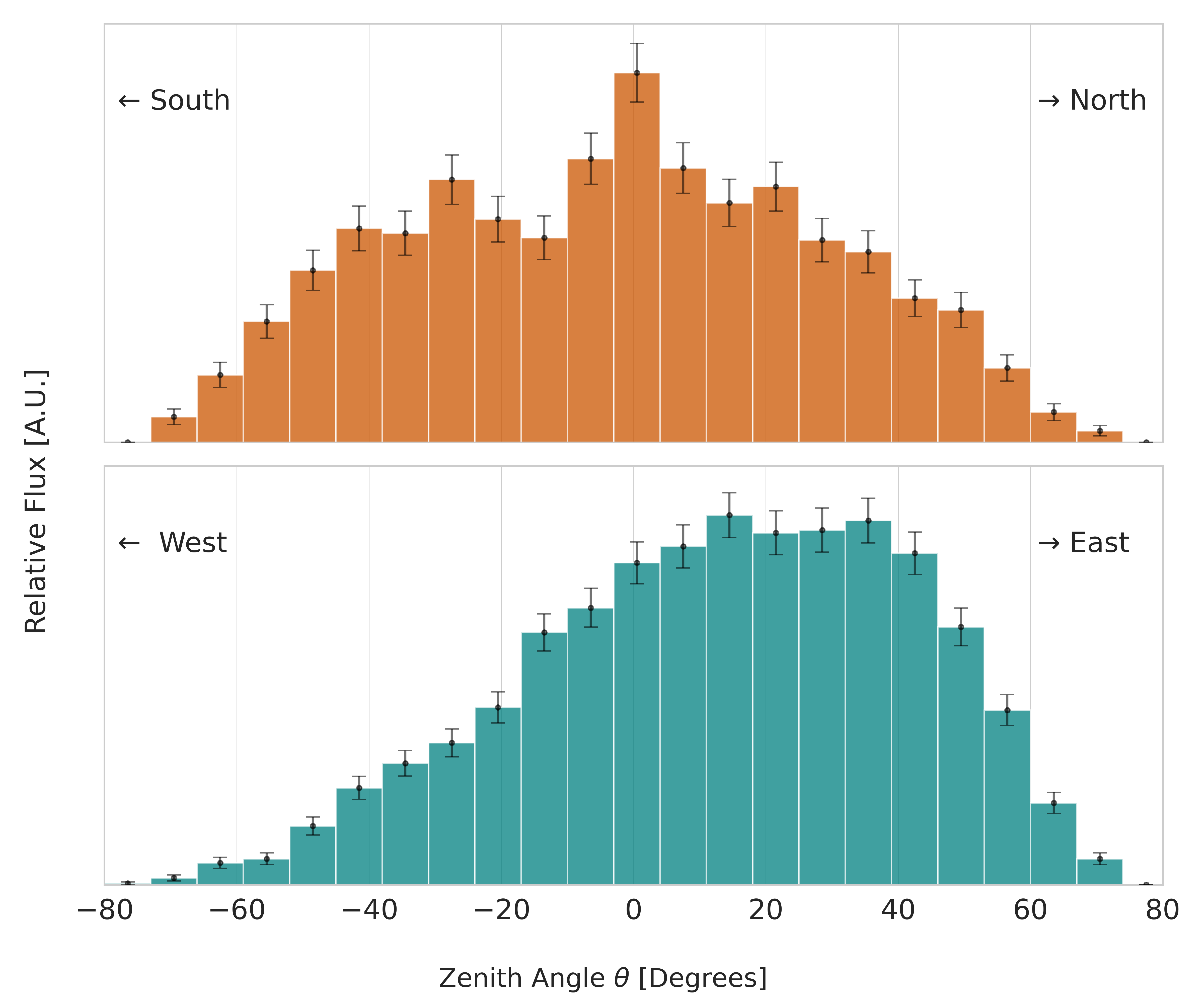}
    \caption{Angular distribution of events from the two orientations at KY. The W-E orientation clearly shows the eastward bias of the muon flux due to the decreasing slope of the mountain profile above in that direction.
    The bin size is of the order of the uncertainty in $\theta$.}
    \label{fig:ang_dist}
\end{figure}

\begin{figure}[h]
    \centering
    \includegraphics[width=.95\linewidth]{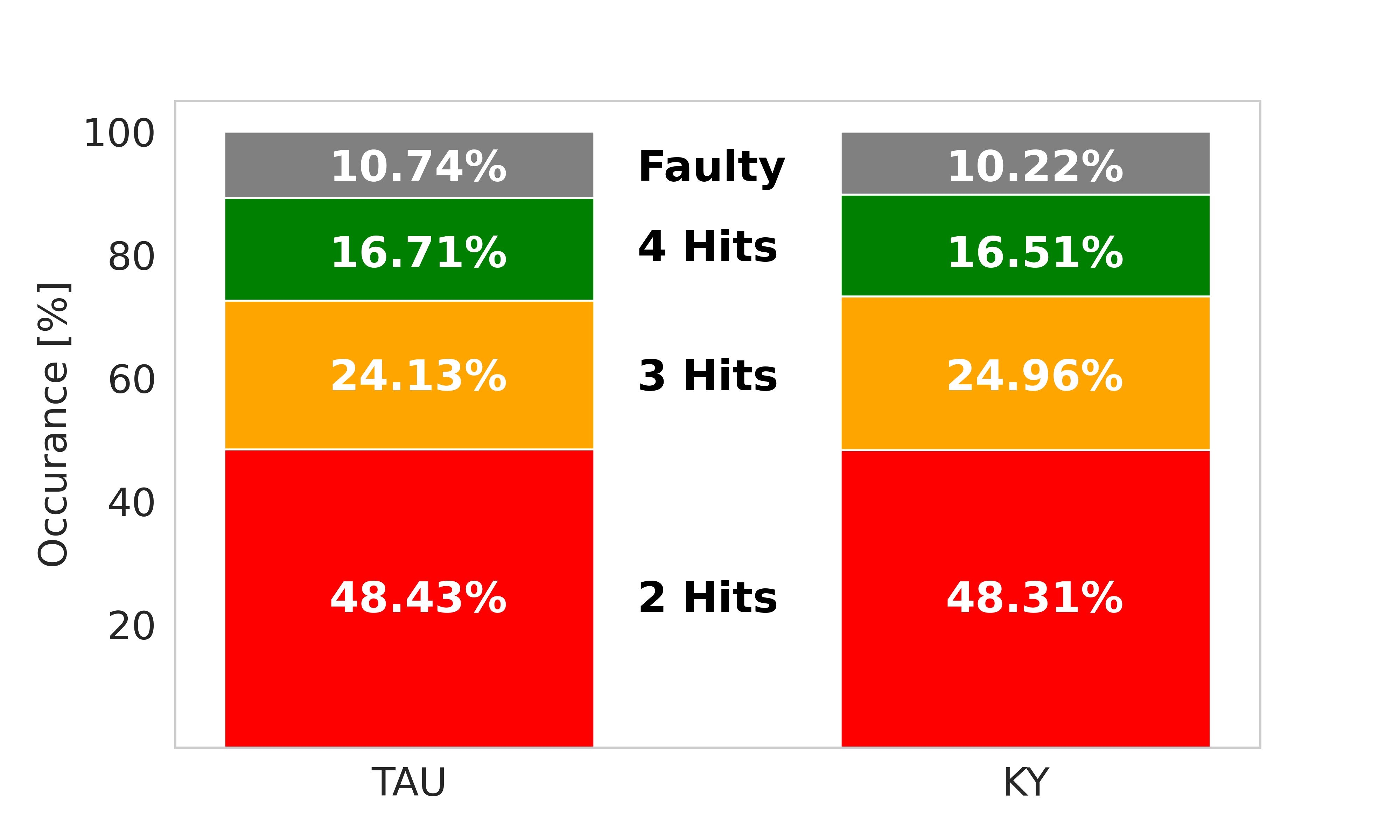}
    \caption{Hit number percentages for each of the campaigns at TAU and KY. Inadequately reconstructed events are denoted ``faulty'' and their zenith angle of arrival $\theta$ is not used.}
    \label{fig:hits_hist}
\end{figure}

We can further see the muon shadow of the mountain in the western direction from the angular distribution of events in figure \ref{fig:ang_dist}, as expected from the increasing slope of the mountain profile. Events are categorised by the number of scintillators they triggered and therefore the number of points used to reconstruct their linear track across the detector. The reconstruction of about 10\% of events in each run failed due to the positioning of some hits outside the detector and were labelled as "faulty" (see figure \ref{fig:hits_hist}). These are subsequently not used in the angular distribution plots in figure \ref{fig:ang_dist}. 

\begin{table}[h]
\centering
\caption{Auxillary measurements of laboratory properties.}\label{aux_tab}%
\begin{tabular}{c|c c c c}
\toprule
Property & Min & Mean & Max  & Units  \\
\midrule
Temperature  & 30.6 & $30.9 \pm 0.2$ & 31.4 & C$^ \circ$ \\
Humidity  & 65.5 & $72.4 \pm 3.0$  & 78.0 & RH\% \\
Rn activity & 11.0 & $28.3 \pm 14.0$ & 76.0 & Bq m$^{-3}$ \\
\botrule
\end{tabular}
\end{table}

Auxiliary measurements of the radon concentrations at KY were also taken using the \textit{Correntium Pro} \cite{airthings2025}, together with additional measurements of the temperature and humidity throughout the measurement campaign and are summarised in table \ref{aux_tab}.

\section{Summary}

The integrated muon flux at the KY underground site was measured to be $3.75 \pm 0.06 \times 10^{-6}$ cm$^{-2}$ s$^{-1}$, a value which translates to a flat overburden of roughly 873 m.w.e..  This puts KY at an intermediate depth to straddle the definitions of shallow and deep underground laboratories. 
Future work will include an effort to commission the space further into the tunnel to win over more overburden. The Rn levels on-site are low; however, the temperature and humidity are consistently high. The working conditions may be dealt with in a future setting as a laboratory. 

The possible uses of a lab with an intermediate overburden may include material screening facilities (e.g.~\cite{Baudis:2011am, vonSivers:2016xpo, Heusser:2015ifa}), testing facilities for delicate components under high voltage and low background, some $\nu$ experiments and low mass Dark Matter (e.g.~\cite{Budnik:2017sbu, SENSEI:2023zdf, DAMIC-M:2023gxo}).  

A separate campaign to measure the residual atmospheric muon flux at the deeper site under Kibbutz Manara will follow with the progress of that site. It could potentially scrape the 2000 m.w.e. mark, confidently placing it in the league with other deep underground laboratories, e.g. WIPP~\cite{Esch:2004zj} which hosted the EXO-200 experiment~\cite{Auger:2012gs}. It of course remains to be seen whether there we will get the sufficient space and working conditions we found at KY.

\section*{Acknowledgments}

This work was supported by the Weizmann Institute of Science Krenter-Perinot Center for High-Energy Particle Physics, the Deloro center and ISF grant 1859/22. 

We thank Gil Doron and Yariv Shalem from Kokhav HaYarden pumped storage facility and Elhanan Hovav for their invaluable assistance and goodwill.

\backmatter

\bibliography{sn-bibliography}

\end{document}